\documentclass[journal]{IEEEtran}
\usepackage{graphicx}
\usepackage{amsfonts}
\usepackage{booktabs}
\usepackage{tabularx} 
\usepackage{multirow}
\usepackage{makecell}
\usepackage{subfigure}
\usepackage{bbding}
\usepackage{bm}
\usepackage{amsmath}
\usepackage{algorithm}
\usepackage{algpseudocode} 
\usepackage{amssymb}
\usepackage{amsfonts}
\usepackage{subcaption}
\usepackage{cite}

\ifCLASSINFOpdf
\else
\fi

\begin{document}
%
\title{Reinforcement Learning Based Symbolic Regression for Load Modeling}
%
%
%

\author{Ding Lin,~\IEEEmembership{Student Member,~IEEE}, Han~Guo,~\IEEEmembership{Student Member,~IEEE}, 
        Jianhui~Wang,~\IEEEmembership{Fellow,~IEEE}, Meng~Yue,~\IEEEmembership{Member,~IEEE}, and Tianqiao~Zhao,~\IEEEmembership{Member,~IEEE}  }
\maketitle

\begin{abstract}
With the increasing penetration of renewable energy sources, growing demand variability, and evolving grid control strategies, accurate and efficient load modeling has become a critical yet challenging task. Traditional methods, such as fixed-form parametric models and data-driven approaches, often struggle to balance accuracy, computational efficiency, and interpretability. This paper introduces a novel symbolic regression algorithm based on the Actor-Critic reinforcement learning framework, specifically tailored for dynamic load modeling. The algorithm employs a trainable expression tree with controlled depth and a predefined set of operators to generate compact and interpretable mathematical expressions. The Actor network  probabilistically selects operators for the symbolic expression, while the Critic evaluates the resulting expression tree through a loss function. To further enhance performance, a candidate pool mechanism is implemented to store high-performing expressions, which are subsequently fine-tuned using gradient descent. By focusing on simplicity and precision, the proposed method significantly reduces computational complexity while preserving interpretability. Experimental results validate its superior performance compared to existing benchmarks, which offers a robust and scalable solution for dynamic load modeling and system analysis in modern power systems.

\end{abstract}


%
\IEEEpeerreviewmaketitle

\section{Introduction}
%
%
%
%
\IEEEPARstart{L}{oad} modeling plays a vital role in power system analysis, particularly with the increasing integration of distributed energy resources and advanced electronic devices in modern power systems. The primary objective of load modeling is to simulate the dynamic response of electrical loads under various disturbances, such as small voltage fluctuations and large-scale grid events. This  process is fundamental to ensuring power system stability, reliability, and operational efficiency. Mathematically, given a dataset $\{x_1,x_2,...,x_n,y\}$, where $\{x_1,x_2,...,x_n\}$ represents measured signals caused by disturbances~(e.g., voltage or frequency variations) and $y$ denotes the expected power response, the goal is to determine an unknown  function $f$ such that $y=f(x_1,x_2,...,x_n)$. With the widespread deployment of high-resolution measurement devices, such as Phasor Measurement Units (PMUs), researchers now have access to abundant data, which enables the development of more accurate load models.

Generally, load modeling methods can be classified into two main categories: fixed-form parametric models and numerical models. Fixed-form parametric models, also known as component-based models, are well-established and widely used. These models are grounded in electromechanical dynamic equations and utilize the physical properties of power system components such as motors, transformers, and electronic devices. They   describe the relationship between load power and variables such as voltage and frequency through mathematical equations. 

Fixed-form parametric models can be  categorized  into static and dynamic load models. Static models, including  exponential models\cite{EM}, constant impedance, current, and power (ZIP) models\cite{ZIP}, and frequency-dependent models\cite{FDM}, describe system behavior without explicitly considering time-dependent dynamics or transient effects.  On the other hand, dynamic models consider the time-dependent response of loads to disturbances. For instance, induction motor (IM) models\cite{ZIP} use equivalent circuit analysis to describe the dynamic variations in active and reactive power, while exponential recovery load models simulate the post-disturbance recovery process using nonlinear first-order equations\cite{ERL1,ERL2}. Furthermore, composite load models (CLM), such as the ZIP+IM\cite{ZIP+IM} model and the Western Electricity Coordinating Council (WECC) CLM\cite{WECC}, combine the features of both static and dynamic models to capture the complex behavior of modern loads under varying operating conditions. While fixed-form parametric models are robust and require minimal measurement data for parameter identification, their application in modern power systems is increasingly challenging due to the growing complexity and nonlinearity of load behavior, which complicates the construction and solution of associated mathematical equations\cite{Linyou}.

To overcome the limitations of fixed-form parametric models, numerical models have emerged as a promising alternative. Unlike their parametric counterparts, numerical models utilize statistical methods and machine learning algorithms to extract load characteristics directly from measurement data. Among these methods, artificial neural networks (ANNs) have been extensively employed for numerical models. For instance, \cite{ANN1} introduced two ANN-based load modeling techniques, while \cite{ANN2} proposed an ANN-based CLM for stability analysis. Furthermore, \cite{LSTM1} employed a multimodal long short-term memory (LSTM) deep learning method to estimate the time-varying parameters of CLM. Similarly, \cite{LSTM2} developed a deep generative architecture based on LSTM networks to identify probabilistic time-varying parameters of dynamic load models. These studies demonstrate the increasing importance of advanced deep learning techniques in capturing the dynamic behavior of complex load systems. However, ANN-based approaches face significant challenges, such as low interpretability and the inability to provide explicit mathematical expressions for load relationships. To address these limitations, symbolic regression (SR) has been introduced as an alternative in numerical load modeling, offering greater interpretability and mathematical clarity.

SR is a data-driven approach designed to automatically discover a mathematical relationship $f$ between $\{x_1,x_2,...,x_n\}$ and $y$ from a given dataset. Unlike traditional regression methods, which rely on predefined functional forms, SR searches the broader space of possible functions to generate explicit mathematical expressions that best describe the data. The final output of SR is typically represented as an expression tree, where internal nodes correspond to mathematical operators and leaf nodes represent variables or constants. This tree structure not only clarifies how input variables are transformed into the output but also provides strong interpretability by revealing the underlying physical or system-specific patterns. Traditional SR methods, often implemented via genetic algorithms or other metaheuristic approaches, offer flexibility but face significant limitations, including low search efficiency, sensitivity to noise, and difficulties in handling high-dimensional and complex problems. Additionally, the expressions generated by these methods are often excessively complex, which reduces interpretability and hinders practical application. Recent advancements in deep learning and reinforcement learning have introduced more sophisticated techniques into SR. Models such as Recurrent Neural Networks (RNNs)\cite{RNNSR} and Transformers\cite{Transformer}, along with reinforcement learning algorithms like policy gradient, have significantly improved the efficiency and accuracy of expression tree construction. However, even these advanced techniques can produce unnecessarily complicated expressions, which limits their interpretability and usability in real-world applications.

To address these challenges, we propose a novel symbolic regression algorithm based on the Actor-Critic reinforcement learning framework. This approach introduces a trainable expression tree structure with controlled depth and a predefined set of operators, which effectively reduces the complexity of generated expressions while preserving interpretability. In this framework, the Actor network probabilistically selects operators for each node in the tree, while the Critic evaluates the tree’s performance using a loss function. By employing a risk-seeking policy gradient strategy, the algorithm prioritizes high-performing expressions, balancing exploration and exploitation during the search process. Additionally, a candidate pool is implemented to store the best-performing expressions identified during training, which are subsequently refined via gradient-based optimization to fine-tune their parameters. This methodology reduces computational overhead while achieving a balance between accuracy and simplicity, which ultimately enhances the efficiency, robustness, and interpretability of symbolic regression models.  Compared to the existing literature, the key contributions of this work are as follows:

\begin{enumerate} 
    \item A symbolic regression algorithm based on the Actor-Critic framework is developed for load modeling. By  leveraging  the Actor to select operators and the Critic to optimize coefficients, the proposed algorithm offers a novel perspective and approach for tackling load modeling challenges. 
    \item We pre-construct a trainable expression tree with a fixed depth and complexity, which effectively mitigates the high computational cost associated with symbolic regression algorithms while simplifying the resulting expression trees. 
    \item A candidate pool is implemented to enhance the optimization process, where expression trees first undergo coarse optimization using policy gradients and are subsequently fine-tuned through gradient descent. Experimental results demonstrate that this approach outperforms existing benchmark methods in load modeling tasks, achieving superior accuracy and efficiency.
\end{enumerate}

The rest of this paper is structured as follows: Section \uppercase\expandafter{\romannumeral2} covers the concepts of symbolic regression based on actor-critic algorithm. Section \uppercase\expandafter{\romannumeral3} provides the case studies. Finally, Section \uppercase\expandafter{\romannumeral4} concludes the paper.

\section{Actor-Critic Based Symbolic Regression}

\begin{figure*}[t] 
    \centering
     \includegraphics[width=\textwidth]{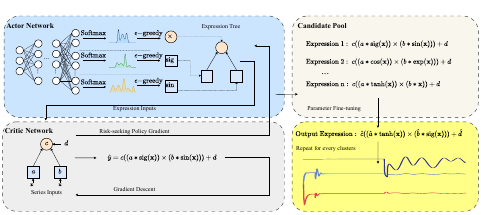} 
    \captionsetup{justification=raggedright, singlelinecheck=false, font=small} 
    \caption{Overview of the proposed actor-critic symbolic regression algorithm.}
    \label{fig: Framework}
\end{figure*}

This section will provide a detailed description of a symbolic regression algorithm based on limiting tree depth and reinforcement learning.

Our algorithm consists of the following steps, illustrated in Fig. \ref{fig: Framework}. First, we limit the complexity of the expressions by defining the depth and structure of a trainable operator expression tree to reduce computational difficulty. Next, we use a Reinforcement Learning approach to probabilistically generate operators for the nodes within the given trainable tree, and apply a greedy algorithm for expression fitting. We then design a reward function tailored to the load modeling task to calculate the reward values and optimize the neural network parameters through gradient descent. Throughout the training process, we maintain a fixed-size candidate pool that stores the best expressions identified. After the neural network training is complete, we further train the trainable tree to obtain the coefficients and biases of the expression.

\subsection{Actor-Critic Deep Reinforcement Learning}
The Actor-Critic algorithm is a widely used method in reinforcement learning that combines the strengths of policy gradient techniques and value function approximation to improve learning efficiency and stability. This algorithm consists of two primary components: the Actor and the Critic.

The Actor network, also known as the policy network, aims to determine a policy that maximizes the expected cumulative reward.  For a given state $s$, the Actor selects an action $a$ with a probability determined by the policy $\pi_{\theta}(a|s)$, where $\theta$ represents the parameters of the policy network. The optimization goal for the Actor is to maximize the expected reward $J(\theta)$, with the gradient given by:
\begin{subequations}
\label{eq: reward function J}
\begin{align}
    J(\theta)=\mathbb{E}_{a \sim \pi_\theta}[A(s, a)] \label{eq1: reward function J} \\
    \theta \leftarrow  \theta  + \alpha \nabla_{\theta} J(\theta),
\end{align}
\end{subequations}
where $A(s, a)$  is the advantage function quantifying how much better an action is compared to the expected value of the state; $\alpha$ denotes the learning rate. Since the goal is to maximize $J$, gradient ascent is used for optimization.

The Critic in the Actor-Critic algorithm is responsible for evaluating the quality of the actions taken by the Actor by estimating the action-value function $Q_w(s,a)$ or the state-value function $V(s)$, depending on the specific variant of the algorithm. The action-value function based Critic is illustrated below.
\begin{subequations}
\label{eq: loss function L}
\begin{align}
    &\mathcal{L}(w) = \frac{1}{N} \sum_{i=1}^N \left( Q_w(s_i, a_i) - \hat{Q} \right)^2 \label{eq1: L(w)}\\
    &w \leftarrow w - \beta \nabla_{w} L(w), \label{eq2: gradient w}
\end{align}
\end{subequations}
where $w$ is the trainable parameters set of the Critic network, $N$ represents the total number of samples in an action batch. $Q$ estimates the expected return of every state-action pair $(s,a)$. $\hat{Q}$ represents the target value used to update the Critic's estimate of the action-value function. The loss function aggregates the errors between the predicted values and the actual returns across all samples. The primary goal of the Critic is to minimize this loss, thereby improving the accuracy of the $Q_w(s_i, a_i)$ estimates. $\beta$ is the learning rate and unlike the training method used for the Actor, the Critic typically optimizes its parameters $w$ using gradient descent.

\begin{figure*}[h] 
    \centering
     \includegraphics[width=\textwidth]{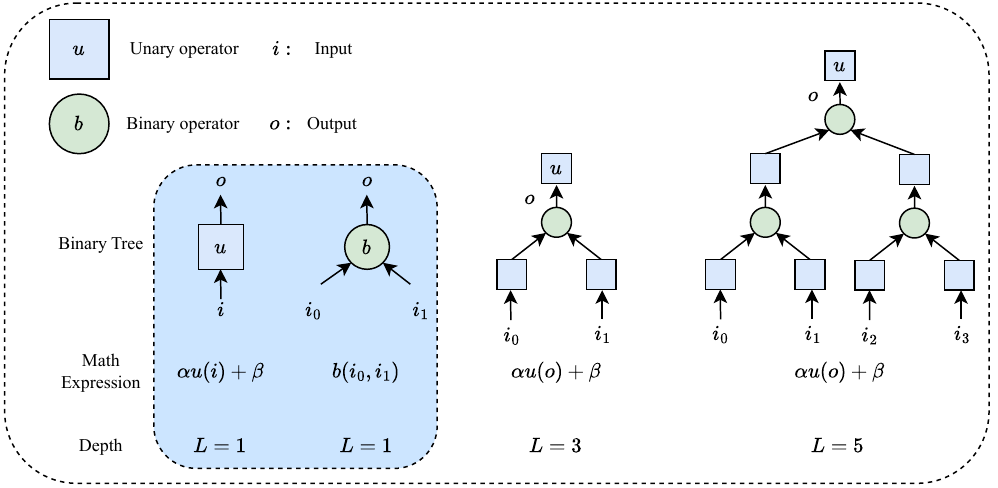} 
    \captionsetup{justification=raggedright, singlelinecheck=false, font=small} 
    \caption{Computational rule of a binary tree.}
    \label{fig: binary tree}
\end{figure*}

\subsection{Trainable Expression Tree}
We introduce a novel trainable expression tree designed specifically for symbolic regression tasks. The structure of this tree is unique in that it alternates between unary and binary operators at each layer, forming a binary tree where the depth of the tree defines the complexity of the expressions it can represent. The definition and the calculation of a binary tree are shown in Fig. \ref{fig: binary tree}.

\subsubsection{Binary Tree Structure}
The trainable expression tree is structured as a binary tree, with each internal node representing either a unary or binary operator, and each leaf node corresponding to an operand or variable.  Unary operators perform operations on a single input, such as activation functions or simple transformations, while binary operators combine two inputs, such as $+$ or $\times$. Each layer of the tree alternates between these two types of operators. In the layers where unary operators are applied, each node processes its input through functions like ReLU, sigmoid, or any other differentiable function, and the resulting output is passed as input to the subsequent binary operator layer. The binary operator layers then combine the outputs of the previous unary operator nodes using operations like addition or multiplication. These operators are selected from our predefined operator library. 

The coefficients for these operators serve as the weights for each node in this binary tree. For instance, the coefficient preceding the operator $\mathrm{exp}$ is the parameter corresponding to that node. These coefficients, along with biases, constitute all the trainable parameters $w$ in the entire trainable tree. Additionally, we divide the selection of operators and the optimization of coefficients into two parts: we use an Actor network to probabilistically select the operators, and a Critic network to optimize the coefficients.

\subsubsection{Parameterization and Training}
For the selection of operators, we use a dedicated Actor to make the selection, and we optimize it using the policy gradient method from reinforcement learning. Here, we will provide a detailed explanation of the parameter optimization within the expression tree. Each operator node in the tree is associated with trainable parameters. For expressions composed of unary operators, such as $a\mathrm{tan}(b\mathbf{x})+c$, these parameters might include coefficients $a,b$ and biases $c$ that scale and shift the input, respectively. For expressions composed of binary operators, such as $a(b\mathbf{x_1}\times c\mathbf{x_2})+d$, the parameters may determine the weighting of the two inputs or other aspects of the operation, depending on the specific operator used.

We treat this expression tree as the Critic. Whenever the Actor samples new operators, these new operators are assigned to the nodes of the expression tree in an in-order traversal manner. We then calculate the output value $\hat{y}$ of the expression based on the operators provided by the Actor and the current 
weights. This paper adopts mean squared error (MSE) as the loss function $M(X, a)$ for the Critic. 
\begin{equation}\label{eq: Tree Loss MSE}
    M(X,a) = \frac{1}{n}\sum_{i=1}^{n}(y_i-\mathcal{T}_w(x_i,a))^2,
\end{equation}
where $w$ refers to all the parameters within the trainable tree. $X$ represents the input to the Critic, which is the sequence of loaded values, $a$ is the set of operators sampled by the Actor, $\mathcal{T}$ is the pre-defined trainable expression tree, and $n$ is the total input samples to the Critic.

Given the MSE function, \eqref{eq1: L(w)} can be rewritten as follows.
\begin{equation}
    \mathcal{L}(w) = \frac{1}{N}\frac{1}{n} \sum_{i=1}^{N} \sum_{j=1}^{n}(y_{j}-\mathcal{T}_w(x_j,a_i))^2,\label{eq: New L(w)}
\end{equation}
where $N$ is the number of actions for one batch. Then, the gradient descent \eqref{eq2: gradient w} optimizes the parameters $w$ of the Critic tree through backpropagation.

\subsection{Policy Gradient}
After the Actor network generates a set of actions $a$ based on a probability, we first optimize the parameters in the Critic tree using this set of actions. Subsequently, we compute the reward function  from the outcomes produced by the trained Critic tree. The Actor network is then trained using this reward function through Policy Gradient methods from reinforcement learning.

\subsubsection{Standard Policy Gradient}
We first examine the standard policy gradient objective, which aims to maximize the expected value of a reward function  under the policy's distribution in our symbolic regression task:
\begin{equation}
J(\theta) = \mathbb{E}_{a \sim \pi_{\theta}} \left[ R(a) \right],
\end{equation}
where $\pi_{\theta}(a)$ is a probability distribution over entire operator sequences, and $R(a)$ is the reward for taking action $a$. The well-known REINFORCE algorithm \cite{williams1992simple} is commonly applied to maximize this expectation using gradient ascent:
\begin{equation}
\begin{split}
    \nabla_{\theta} J(\theta) &= \nabla_{\theta}\mathbb{E}_{a \sim \pi_{\theta}} \left[ R(a) \right] \\
    &= \mathbb{E}_{a \sim \pi_{\theta}} \left[ R(a) \nabla_{\theta}\mathrm{log}\pi_{\theta}(a) \right].
\end{split}
\end{equation}

This approach enables the approximation of the expectation by sampling from the distribution. Specifically, an unbiased estimate of $\nabla_{\theta} J(\theta)$ can be obtained by computing the average gradient over a batch of $N$ sampled trajectories $\{a^{(i)}\}_{i=1}^{N}$:
\begin{equation}
    \nabla_{\theta} J(\theta) \approx \frac{1}{N}\sum_{i=1}^{N}R(a^{(i)})\nabla_{\theta}\mathrm{log}\pi_{\theta}(a^{(i)}).
\end{equation}

While this gradient estimate is unbiased, it often suffers from high variance in practice. To mitigate this variance, it is common practice to subtract a baseline function $b$ from the reward $R(a)$. Provided that this baseline does not depend on the specific trajectory $a$ of the current batch, the gradient estimate remains unbiased. Typical choices for the baseline include moving averages of the rewards or estimates of the value function. Conceptually, the gradient step increases the probability of trajectories that yield rewards above the baseline while decreasing the probability of those below it.

\subsubsection{Risk-Seeking Policy Gradient}
The standard policy gradient objective, $J(\theta)$, is typically framed as an expectation, which is not always suitable for problems where the goal is to maximize the top-end performance rather than the average. For example, in tasks like symbolic regression, program synthesis, or neural architecture search, final performance is often measured based on the best samples observed during training. Similarly, one might be more interested in policies that achieve "high scores" in control environments. However, for such scenarios, $J(\theta)$ may not be ideal as it focuses on the average, which potentially leads to suboptimal results when aiming to optimize the highest possible outcomes. 

To address this, we utilize the risk-seeking policy gradient in \cite{RNNSR}. This policy defines $R_{\epsilon}(a)$ as the $(1-\epsilon)$-quantile of the distribution of rewards under the current policy, which is not a function of $\theta$ but is typically intractable. Thus a new learning objective, $J_{\mathrm{rish}}(\theta;\epsilon)$ is introduced:
\begin{equation}
    J_{\mathrm{rish}}(\theta;\epsilon) = \mathbb{E}_{a \sim \pi_{\theta}}\left[R(a)\mid R(a)\geq R_{\epsilon}(\theta) \right].
\end{equation}

This objective aims to increase the reward for the top $\epsilon$ fraction of samples, without regard for samples below this threshold. The gradient of $J_{\mathrm{rish}}(\theta;\epsilon)$ is then expressed as:
\begin{equation}
\begin{split}
    & \nabla_{\theta} J_{\text{risk}}(\theta; \epsilon) = \mathbb{E}_{a \sim \pi_{\theta}} \\ 
    & \left[ (R(a) - R_{\epsilon}(\theta)) \cdot \nabla_{\theta} \log \pi_{\theta} 
    \mid R(a) \geq R_{\epsilon}(\theta) \right].
\end{split} 
\end{equation}
This result offers a straightforward Monte Carlo estimate of the gradient from a batch of $N$ samples:
\begin{equation}\label{eq: J_risk}
\begin{split}
    & \nabla_{\theta} J_{\text{risk}}(\theta; \epsilon) \approx \frac{1}{N}\sum_{i=1}^{N} \\ 
    & \left[ (R(a^{(i)}) - R_{\epsilon}(\theta)) \cdot \nabla_{\theta} \log \pi_{\theta}(a^{(i)}) 
    \cdot \mathbb{I}_{R(a^{(i)}) \geq \hat{R}_{\epsilon}(\theta)} \right],
\end{split} 
\end{equation}
where $\hat{R}_{\epsilon}(\theta)$ is the empirical $(1-\epsilon)$-quantile of the rewards from the batch. The term $\mathbb{I}$ is an indicator function that equals 1 if the condition is satisfied and 0 otherwise.

The risk-seeking policy is similar to the standard REINFORCE Monte Carlo estimate, but the rewards are adjusted by subtracting the baseline $\hat{R}_{\epsilon}(\theta)$, which is a proxy for the standard policy gradient baseline. Both two policies will be compared in the case study section.

Given a dataset $\{X,y\}_{i=1}^{n}$, and a Critic tree function $f$. To bound the reward function, we apply a squashing function using Root Mean Squared Error (RMSE):
\begin{equation}\label{eq: Advantage function RMSE}
    R(a) = \frac{1}{1+\sqrt{\frac{1}{n}\sum_{i=1}^{n}(y_{i}-f(X_{i}))^{2}}}.
\end{equation}

\subsection{Candidate Pool and Fine-tuning}
Generally, the optimization based on \eqref{eq: New L(w)} is typically non-convex, so the parameters of the Critic Tree obtained through coarse-tuning may be prone to local optima. To address this issue, we store the actions with reward values ranked in the top $\mathcal{C}$ during the training process in the candidate pool $\mathcal{P}$. 
After the Actor-Critic training is completed, we will fine-tune the expression trees in $\mathcal{P}$ using first-order optimization, similar to the method in \eqref{eq2: gradient w}. Once all the expressions in the pool have been fine-tuned, we will select the expression with the smallest loss as the result of the symbolic regression of the input dataset $\{X,y\}_{i=1}^{n}$.

\begin{algorithm}[!h]\label{alg:sampling}
  \caption{Actor-critic based symbolic regression for load modeling} 
  \begin{algorithmic}[1]
  \Require The load data $\{X,y\}_{i=1}^{n}$ and the pre-defined function; An actor network $\mathcal{A}$ and a critic tree $\mathcal{T}$; Training iterations for actor $I_1$ and critic $I_2$; Fine-tuning iteration $I_3$; Candidate pool size $\mathcal{C}$ and action batch size $N$.
  \Ensure Expression $f$ based on the critic tree $\mathcal{T}$.
    \For{$i=1,...,I_1$}
        \State Sample $N$ operator sequences $\{a^{(1)},...,a^{(N)}\}$ from $\mathcal{A}$.
        \For{$j=1,...,N$}
            \State Calculate MSE function $M(X,a^{(j)})$ according to \eqref{eq: Tree Loss MSE}.
            \For{$k=1,...,I_2$}
                \State Train the Critic tree $\mathcal{T}$ using \eqref{eq2: gradient w}.
            \EndFor
            \State  Compute the reward function $R$ based on \eqref{eq: Advantage function RMSE}, and update the candidate pool $\mathcal{P}$.
            \State Updating $\mathcal{A}$ according to \eqref{eq: J_risk}.
        \EndFor 
    \EndFor
    \For{$a \in \mathcal{C}$}
        \For{$i=1,...,I_3$}
            \State Calculate $M(X,a)$, and fine-tuning $\mathcal{T}$ using \eqref{eq2: gradient w}.
        \EndFor
        \State Save $a$ with minimum $M(X,a)$.
    \EndFor
    \State \textbf{Return} The expression $f$ generated according to in-order traversal of the fine-tuning tree $\mathcal{T}$, with the minimum $M(X,a)$.
  \end{algorithmic}
\end{algorithm}

\section{Case Study}
\subsection{Dataset Description}

\begin{table*}[h]
\centering
\renewcommand{\arraystretch}{1.3}
\setlength{\tabcolsep}{7pt}
\caption{The RMSE Results For Different Risk-seeking Policies}
\label{table: Risk-seeking policy}

\begin{tabularx}{\textwidth}{c|l|X|X|X|X|X|X|X|X|X|X|X|X}
\toprule

\multirow{2}{*}{\textbf{Fault Type}} & \multirow{2}{*}{\textbf{Load Type}} & \multicolumn{3}{c|}{\textbf{Risk Policy $\epsilon=0.3$}} & \multicolumn{3}{c|}{\textbf{Risk Policy $\epsilon=0.5$}} & \multicolumn{3}{c|}{\textbf{Risk Policy $\epsilon=0.7$}} & \multicolumn{3}{c}{\textbf{Standard Policy}} \\
\cmidrule{3-14}
 &  & \textbf{V} & \textbf{P} & \textbf{Q}  & \textbf{V} & \textbf{P} & \textbf{Q} & \textbf{V} & \textbf{P} & \textbf{Q}  & \textbf{V} & \textbf{P} & \textbf{Q} \\
\midrule
\multirow{3}{*}{Bus Fault}  & CI  & 0.063 & 0.070 & 0.084 & \textbf{0.055} & \textbf{0.070} & \textbf{0.083} & 0.098 & 0.071 & 0.083 & 0.077 & 0.070 & 0.083 \\
                            & IM & 0.046 & \textbf{0.055} & 0.016 & 0.044 & 0.059 & \textbf{0.014} & \textbf{0.040} & 0.056 & 0.015  & 0.044 & 0.057 & 0.015\\
                            & PV & \textbf{0.041} & 0.034 & 0.026 & 0.042 & 0.034 & \textbf{0.025} & 0.042 & \textbf{0.032} & 0.026  & 0.047 & 0.035 & 0.026\\
\midrule
\multirow{3}{*}{Line Fault}  & CI  & 0.045 & 0.071 & 0.085 & \textbf{0.043} & \textbf{0.069} & 0.085 & 0.045 & 0.068 & 0.086 & 0.046 & 0.071 &  \textbf{0.083} \\
                            & IM & 0.041 &  0.019 & 0.059 & \textbf{0.039} & \textbf{0.019} & \textbf{0.057} & 0.040 & 0.019 & 0.057  &  0.041 & 0.020 & 0.057\\
                            & PV &  0.042 & 0.041 & 0.023 & \textbf{0.040} & \textbf{0.039} & \textbf{0.023} & 0.045 & 0.039 & 0.023  & 0.049 & 0.040 & 0.023\\

\bottomrule
\end{tabularx}
\end{table*}

\begin{table*}[h]
\centering
\renewcommand{\arraystretch}{1.3}
\setlength{\tabcolsep}{7pt}  
\caption{The RMSE Results For Different Depths}
\label{table: tree depth}

\begin{tabularx}{\textwidth}{c|l|X|X|X|X|X|X|X|X|X}
\toprule

\multirow{2}{*}{\textbf{Fault Type}} & \multirow{2}{*}{\textbf{Load Type}} & \multicolumn{3}{c|}{\textbf{$L=3$}} & \multicolumn{3}{c|}{\textbf{$L=5$}} & \multicolumn{3}{c}{\textbf{$L=7$}}  \\
\cmidrule{3-11}
 &  & \textbf{V} & \textbf{P} & \textbf{Q}  & \textbf{V} & \textbf{P} & \textbf{Q} & \textbf{V} & \textbf{P} & \textbf{Q}  \\
\midrule
\multirow{3}{*}{Bus Fault}  & CI  & 0.063 & 0.078 & 0.090 & 0.055 & \textbf{0.070} & 0.083 & \textbf{0.053} & 0.072 & \textbf{0.082}  \\
                            & IM & \textbf{0.040} & 0.067 & 0.031 & 0.044 & \textbf{0.059} & 0.014 & 0.047 & 0.060 & \textbf{0.011} \\
                            & PV & 0.049 & 0.040 & 0.033 & \textbf{0.042} & 0.034 & \textbf{0.025} & 0.044 & \textbf{0.032} & 0.027 \\
\midrule
\multirow{3}{*}{Line Fault}  & CI  & 0.051 & 0.076 & 0.094 & 0.043 & \textbf{0.069} & \textbf{0.085} & \textbf{0.039} & 0.071 & 0.087\\
                            & IM & 0.050 & 0.037 & 0.065 & 0.039 & \textbf{0.019} & \textbf{0.057} & \textbf{0.037} & 0.025 & 0.059  \\
                            & PV & 0.046 & \textbf{0.038} & 0.036 & 0.040 & 0.039 & \textbf{0.023} & \textbf{0.039} & 0.042 & 0.025\\

\bottomrule
\end{tabularx}
\end{table*}

\begin{table*}[h]
\centering
\renewcommand{\arraystretch}{1.3}
\setlength{\tabcolsep}{7pt}  
\caption{The RMSE For Different Models}
\label{table: different models}

\begin{tabularx}{\textwidth}{c|l|X|X|X|X|X|X|X|X|X|X|X|X}
\toprule

\multirow{2}{*}{\textbf{Fault Type}} & \multirow{2}{*}{\textbf{Load Type}} & \multicolumn{3}{c|}{Proposed Model} & \multicolumn{3}{c|}{ANN}& \multicolumn{3}{c|}{ZIP} & \multicolumn{3}{c}{Polynomial}  \\
\cmidrule{3-14}
 &  & \textbf{V} & \textbf{P} & \textbf{Q}  & \textbf{V} & \textbf{P} & \textbf{Q} & \textbf{V} & \textbf{P} & \textbf{Q} & \textbf{V} & \textbf{P} & \textbf{Q}  \\
\midrule
\multirow{3}{*}{Bus Fault}  & CI  & 0.055 & \textbf{0.070} & \textbf{0.083} & \textbf{0.051} & 0.078 & 0.092 &  0.652 & 0.783&0.943& 2.188 & 3.068 & 5.394 \\
                            & IM & 0.044 & \textbf{0.059} & \textbf{0.014} & \textbf{0.040} & 0.066 & 0.012 & 0.536 & 0.694  & 0.253 & 1.759 & 2.579 & 2.629\\
                            & PV & \textbf{0.042} & \textbf{0.034} & \textbf{0.025} & 0.047 & 0.036 & 0.029 & 0.579 & 0.461 & 0.382& 1.427 & 1.385 & 3.173  \\
\midrule
\multirow{3}{*}{Line Fault}  & CI  & \textbf{0.043} &  \textbf{0.062}& \textbf{0.085} & 0.047 & 0.069 & 0.088 & 0.428 & 0.727 & 0.983 & 1.526 & 3.577 & 5.851\\
                            & IM & \textbf{0.039} & \textbf{0.019} & \textbf{0.057} & 0.042 & 0.021 & 0.066 & 0.403 & 0.279 & 0.596& 1.339 & 1.231 & 3.358  \\
                            & PV & \textbf{0.040} & 0.039 & \textbf{0.023} & 0.050 & \textbf{0.033} & 0.028 & 0.416 & 0.529 & 0.296& 1.735 & 1.636 & 2.729\\

\bottomrule
\end{tabularx}
\end{table*}

The dataset used in this study is based on simulations performed on a modified IEEE 39-bus system and a 33-node feeder model. Disturbances are introduced at 3.5 seconds, and data is recorded under both bus fault and line fault scenarios. The load types in the system include constant impedance (CI), induction motors (IM), and photovoltaic distributed energy resources (PV-DER). Specifically, the system consists of 10 CIs, 3 IMs, and 4 PVs, among other load types. For each fault scenario, one CI load, one IM load, and one PV-DER unit are selected to showcase the experimental results. Key measurements such as voltage, active power, and reactive power (V, P, Q) are captured at these load points, providing a comprehensive dataset for dynamic load modeling. This setup ensures the proposed algorithm is rigorously evaluated across diverse load types and fault conditions, which highlights its robustness and adaptability.

\begin{figure*}[h] 
    \centering
     \includegraphics[width=\textwidth]{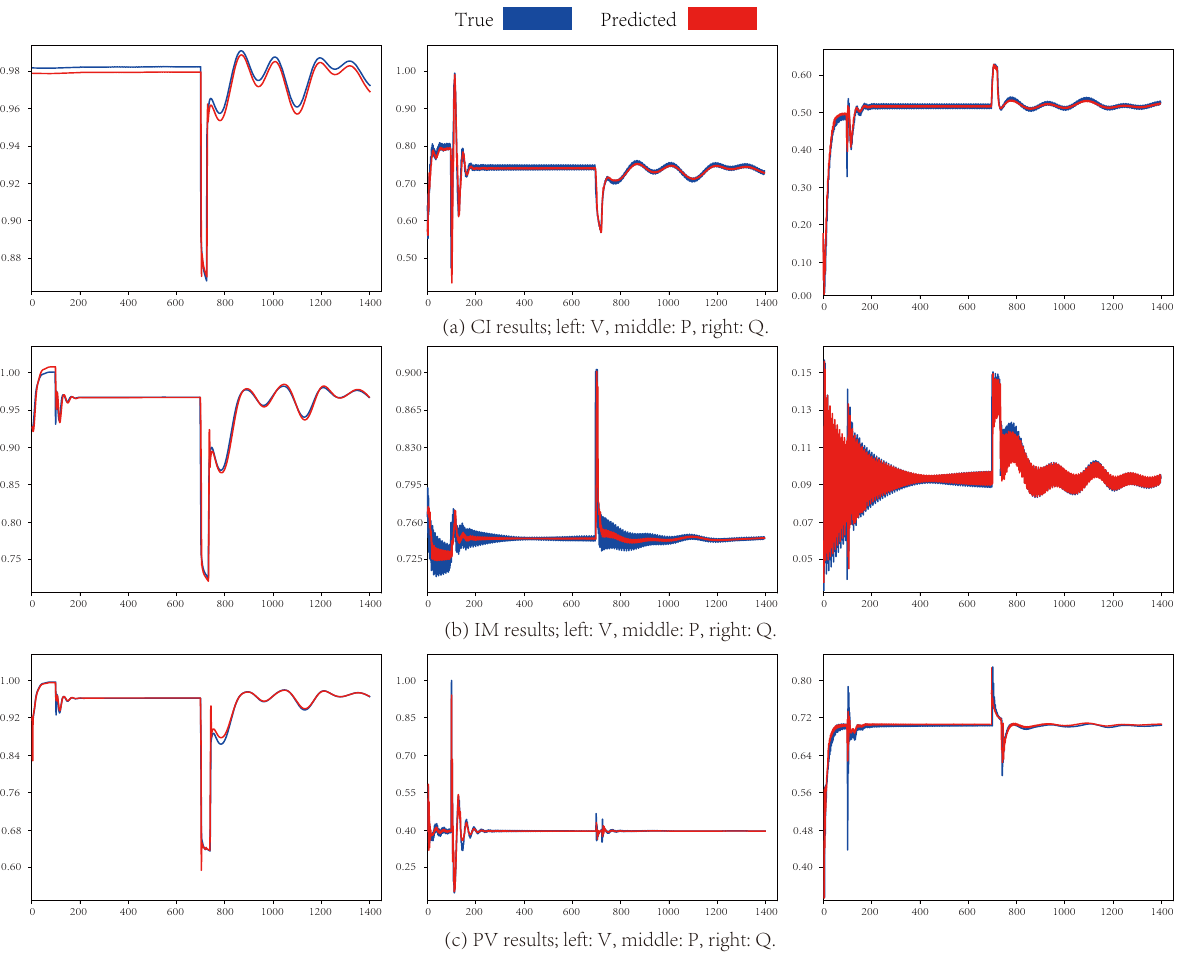} 
    \captionsetup{justification=raggedright, singlelinecheck=false, font=small} 
    \caption{Performance evaluation of the proposed symbolic regression algorithm under bus fault scenario.}
    \label{fig: bus fault}
\end{figure*}

\begin{figure*}[h] 
    \centering
     \includegraphics[width=\textwidth]{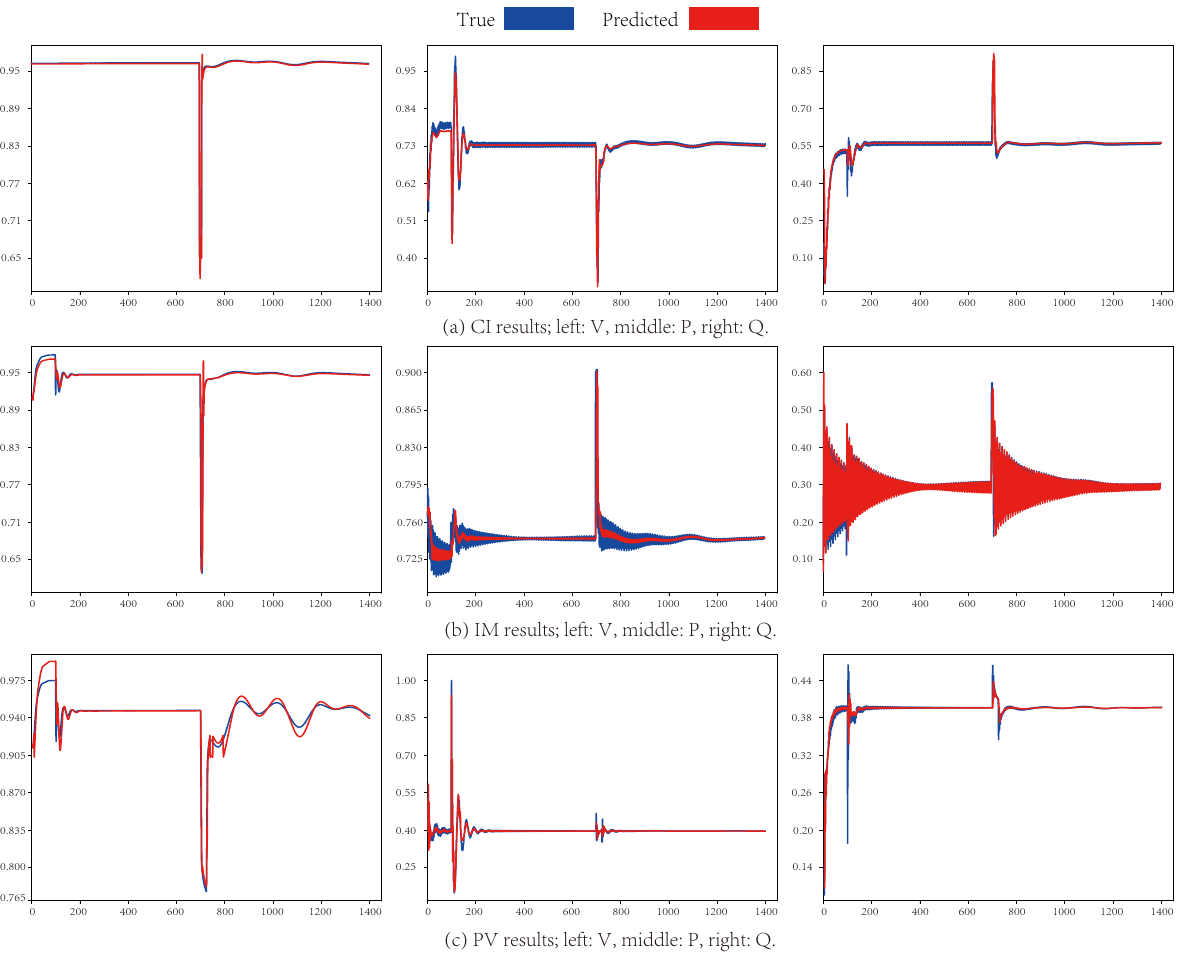} 
    \captionsetup{justification=raggedright, singlelinecheck=false, font=small} 
    \caption{Performance evaluation of the proposed symbolic regression algorithm under line fault scenario.}
    \label{fig: line fault}
\end{figure*}

\subsection{Selection of Risk-Seeking Policy}

The selection of the risk-seeking policy parameter \(\epsilon\) in reinforcement learning is critical for balancing the trade-off between emphasizing extreme samples and maintaining overall model accuracy. To evaluate the impact of different risk-seeking policies, the model's performance was tested under varying \(\epsilon\) values (\(\epsilon = 0.3, 0.5, 0.7\)) and compared with the standard policy. Root Mean Square Error (RMSE) was used as the evaluation metric, with results summarized in Table \ref{table: Risk-seeking policy} across bus fault and line fault scenarios for three types of loads.

The results indicate that \(\epsilon=0.5\) consistently achieves the best overall performance across various load types and fault scenarios. For instance, in the bus fault scenario with CI loads, the RMSE for $V$ improves significantly from 0.063 at \(\epsilon=0.3\) to 0.055 at \(\epsilon=0.5\). However, at \(\epsilon=0.7\), the RMSE increases to 0.098, suggesting that excessive risk-seeking behavior can degrade performance by over-prioritizing extreme samples at the expense of overall stability. A similar pattern is observed in the line fault scenario. For example, with PV loads, the RMSE for $V$ decreases from 0.042 at \(\epsilon=0.3\) to 0.040 at \(\epsilon=0.5\). When \(\epsilon=0.7\), the RMSE slightly increases to 0.045, which further supporting that overly aggressive risk-seeking behavior compromises the model's generalization ability. Across all load types, \(\epsilon=0.5\) strikes an optimal balance and allows the model to prioritize high-quality expressions while maintaining accuracy and robustness. This balanced performance is evident in both $P$ and $Q$ predictions. For example, in the line fault scenario with IM loads, \(\epsilon=0.5\) achieves an RMSE of 0.019 for $Q$, outperforming both \(\epsilon=0.3\) and \(\epsilon=0.7\). These findings highlight that \(\epsilon=0.5\) not only ensures reliable performance across a wide range of conditions but also avoids the pitfalls of both overly conservative and overly aggressive risk-seeking strategies.

Based on these results, \(\epsilon = 0.5\) is selected as the optimal risk-seeking policy for subsequent evaluations. This choice ensures a stable and reliable prediction framework, and optimizes the trade-off between focusing on top-performing samples and maintaining overall accuracy.

\subsection{Selection of Tree Depth}

The depth of the binary expression tree \(L\) is another crucial factor in the proposed symbolic regression model, which  directly influences its complexity, expressiveness, and computational efficiency. To evaluate the impact of different tree depths, the model's performance was tested under varying depths (\(L = 3, 5, 7\)) across bus fault and line fault scenarios, considering three load types: CI, IM and PV. The RMSE results are summarized in Table \ref{table: tree depth}.

The results consistently demonstrate that a tree depth of \(L = 5\) achieves the best overall performance across all load types and fault scenarios. For example, in the bus fault scenario with CI loads, the RMSE for $V$ decreases from 0.063 at \(L = 3\) to 0.055 at \(L = 5\), which reflects the improved ability of the model to capture more complex patterns. However, increasing the tree depth further to \(L = 7\) results in a slight increase in RMSE to 0.053, indicating that excessive complexity may introduce overfitting without significant performance gains.

Similar trends are observed in the line fault scenario. For instance, with PV loads, the RMSE for $V$ decreases from 0.046 at \(L = 3\) to 0.040 at \(L = 5\), but further increasing the depth to \(L = 7\) offers minimal improvement, with an RMSE of 0.039. This demonstrates that while deeper trees can potentially enhance expressiveness, they also lead to increased computational overhead and the risk of overfitting. For $P$ and $Q$ metrics, the results further validate $L=5$ as the optimal choice. In the line fault scenario with IM loads, for example, the RMSE for $Q$ improves significantly from 0.065 at \(L = 3\) to 0.057 at \(L = 5\), but increases slightly to 0.059 at $L=7$. This indicates $L=5$ strikes an ideal balance between complexity and accuracy.

Overall, the results show that a tree depth of $L=5$ is optimal for the proposed model. It provides sufficient capacity to capture the nonlinear relationships in load modeling tasks while avoiding the risks of overfitting and excessive computational costs associated with deeper trees. Based on these findings, $L=5$ is selected as the tree depth for subsequent experiments, which ensures an effective trade-off between accuracy, complexity, and computational efficiency.

\subsection{Comparative Experiment}

To validate the effectiveness of the proposed symbolic regression model, its performance was compared with three baseline models: ANN, ZIP, and Polynomial Regression. The RMSE results for all four methods under different fault scenarios and load types are summarized in Table \ref{table: different models}.

The results demonstrate that the proposed model consistently outperforms the baseline methods in most cases, achieving lower RMSE values across fault types and load conditions. For instance, in the bus fault scenario with CI loads, the proposed model achieves an RMSE of 0.055 for $V$, which is slightly higher than ANN but significantly lower than ZIP and Polynomial Regression. Similarly, for $P$ and $Q$ under the same scenario, the proposed model delivers the best performance with an RMSE of 0.070 and 0.083, respectively. In the line fault scenario, the superiority of the proposed model is even more evident. For PV loads, the RMSE for voltage decreases to 0.040 with the proposed model, outperforming all the baseline models. Additionally, for $Q$ in the same scenario, the proposed model achieves the lowest RMSE of 0.023, which further highlights its effectiveness.

Across all load types, the proposed model demonstrates a balanced performance, achieving the best or near-best results for $V$, $P$, and $Q$. Notably, while ANN occasionally delivers comparable results for specific metrics, such as $V$, its performance is less consistent across all fault scenarios and load conditions. ZIP and Polynomial Regression, on the other hand, consistently lag behind both the proposed model and ANN, which highlights their limited capability to capture the nonlinear relationships inherent in load modeling tasks. These results validate the effectiveness of the proposed symbolic regression model, which achieves superior accuracy while maintaining interpretability and computational efficiency. Its ability to outperform ANN, ZIP, and Polynomial Regression in most scenarios underscores its robustness and adaptability for dynamic load modeling tasks.

\subsection{Performance Evaluation of Two Fault Scenarios}

Fig. \ref{fig: bus fault} and \ref{fig: line fault} illustrate the performance of the proposed symbolic regression algorithm under bus fault and line fault scenarios, focusing on three types of loads: CI, IM, and PV. For each load type, the algorithm's predictions for $V$, $P$, and $Q$ are compared against the true values, with predicted values shown in red and true values in blue.
Across both fault scenarios, the algorithm consistently demonstrates strong performance in modeling the dynamic responses of all three load types. Predictions for $V$ show a high level of accuracy across all load types, with predicted values closely matching true values even during transient periods following the faults. This highlights the robustness of the algorithm in capturing both steady-state and transient voltage behaviors.
Similarly, predictions for $P$ exhibit strong agreement with true values, with only minor deviations observed during transient responses, particularly for IM loads, which display more complex dynamic characteristics.
Predictions for $Q$ are also highly accurate for CI and PV loads, while slightly larger deviations are observed for IM loads during fault transients, suggesting potential areas for further optimization.

The excellent performance across both fault scenarios  showcases the algorithm's adaptability and robustness in addressing diverse fault conditions. Its ability to effectively capture the dynamic behavior of modern power systems, including PV loads, which are known for their variability and nonlinearity, further underscores its practical utility. Although minor deviations are observed in the IM load predictions for $P$ and $Q$ under transient conditions, the results validate the proposed method's effectiveness in accurately modeling dynamic load behavior while maintaining interpretability. These findings highlight the algorithm's potential for practical applications in load modeling, which offers a scalable and robust solution for analyzing complex fault scenarios in modern power systems.

\section{Conclusion}

This paper proposed a symbolic regression algorithm based on the Actor-Critic reinforcement learning framework for load modeling. By introducing a trainable expression tree with controlled depth and complexity, the algorithm effectively balances efficiency and interpretability. The Actor network probabilistically selects operators for expression tree, while the Critic evaluates it using a loss function based on MSE. To further improve accuracy and robustness, a candidate pool mechanism is employed to refine the best-performing expressions. Experiments conducted under bus fault and line fault scenarios demonstrated the algorithm's capability to model various load types, including constant impedance, induction motors, and photovoltaic distributed energy resources. The proposed method achieved high accuracy in predicting voltage, active power, and reactive power, while outperforming benchmarks. This study provides an efficient, robust, and interpretable solution for load modeling, offering a novel approach to capturing complex dynamic behaviors in modern power systems.

\ifCLASSOPTIONcaptionsoff
  \newpage
\fi

\bibliographystyle{IEEEtran}
\bibliography{IEEEabrv, ref.bib}

\end{document}